\documentclass[journal]{IEEEtran}
\usepackage{cite}
\usepackage{amsmath,amssymb,amsfonts}
\usepackage{amsthm}
\usepackage{nccmath}
\usepackage{fancyhdr}
\usepackage{svg}
\usepackage{bbm}
\usepackage{soul}
\usepackage{url}
\usepackage{subfloat}
\usepackage{graphicx}
\usepackage{textcomp}
\usepackage{xcolor}
\usepackage{kotex}
\usepackage{algorithmicx,algpseudocode}
\usepackage{subfigure}
\usepackage{booktabs}
\usepackage{soul,color}
\usepackage{bbold}
\usepackage{mwe}
\usepackage{multicol}
\usepackage{multirow}
\usepackage[normalem]{ulem}
\usepackage[linesnumbered,ruled]{algorithm2e}

\newcommand{\rev}[1]{{\color{black}{#1}}}
\newcommand{\BfPara}[1]{{\noindent\bf#1.}\xspace}

\def\BibTeX{{\rm B\kern-.05em{\sc i\kern-.025em b}\kern-.08em
    T\kern-.1667em\lower.7ex\hbox{E}\kern-.125emX}}

\begin{document}

\title{\rev{Quantum Neural Network Software Testing, Analysis, and Code Optimization for Advanced IoT Systems}: Design, Implementation, and Visualization}


\author{
    Soohyun Park 
    and Joongheon Kim,~\IEEEmembership{Senior Member, IEEE}  

    \thanks{S. Park and J. Kim are with the School of Electrical Engineering, Korea University, Seoul 02841, Republic of Korea (e-mails: \{soohyun828,joongheon\}@korea.ac.kr).}
}
\maketitle

\begin{abstract}
\rev{This paper introduces a novel run-time testing, analysis, and code optimization (TACO) method for quantum neural network (QNN) software in advanced Internet-of-Things (IoT) systems, which visually presents the learning performance that is called a barren plateau.} 
The \rev{run-time} visual presentation of barren plateau situations is helpful for \rev{real-time} quantum-based  \rev{advanced IoT} software testing because the software engineers can easily be aware of the training performances \rev{of QNN}. Moreover, this tool is obviously useful for software engineers because it can intuitively guide them in designing and implementing high-accurate \rev{QNN-based advanced IoT} software even if they are not familiar with quantum mechanics and quantum computing. Lastly, the proposed TACO is also capable of visual feedback because software engineers visually identify the barren plateau situations using tensorboard. In turn, they are also able to modify \rev{QNN} structures based on the information.
\end{abstract}
\section{Introduction}\label{sec:intro}

\begin{figure}[t]
\centering\includegraphics[width=0.99\columnwidth]{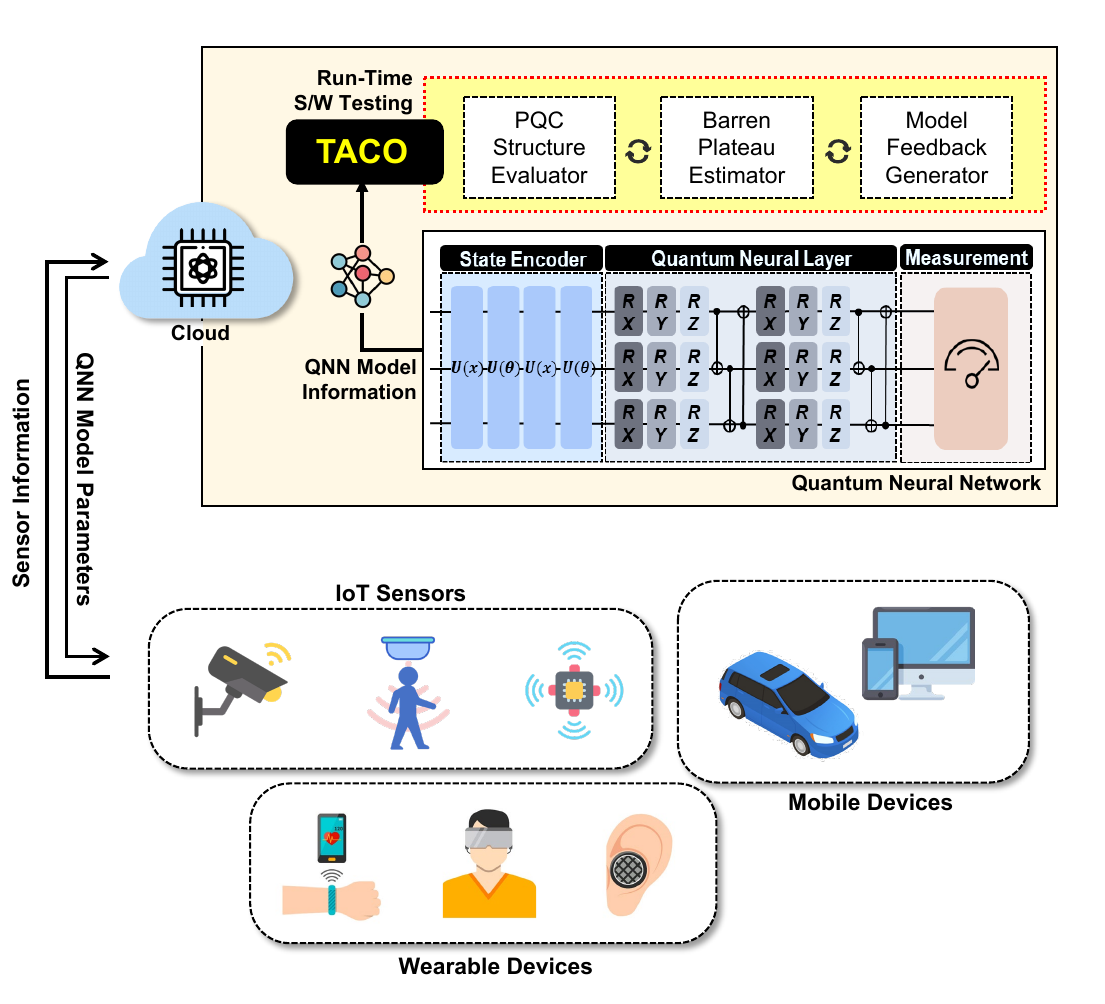}
    \caption{Run-Time Software Testing \rev{and Analysis} in \rev{QNN}-based IoT Systems.}
    \label{fig:sec2}
\end{figure}

\rev{In modern daily lives, every single person actively starts to utilize Internet services not only in smartphones but also in wearable devices such as smartwatches and smartglasses. This trend can be realized due to the advances in Internet-of-Things (IoT) technologies where the ultimate mission of IoT is for enabling global Internet connection among all IoT devices which can realize the utilization of the perceived data from built-in sensors in the devices. For the example of IoT-based autonomous driving, the} technologies have undergone significant development, and recent advances have made precise low-delay real-time perception of \rev{surrounding} driving environments~\cite{ic2309park}. As \rev{presented} in Fig.~\ref{fig:sec2}, the environment perception involves recognition objects from \rev{various} sensors \rev{using wireless} communications, which deliver sensor information from \rev{IoT devices} to \rev{their associated centralized server}~\cite{tvt2108jung}. 
\rev{After gathering sensor information from deployed IoT devices, the server conducts deep learning computation for various application-specific AI services such as detection, classification, and recognition. Then, the trained parameters will be delivered back to the individual IoT devices in order to realize the services. In this scenario, the use of quantum neural network (QNN)-based deep learning models is definitely beneficial for computation/memory-limited IoT devices due to the fact that QNN-based models can utilize much fewer parameters comparing to conventional DNN-based models~\cite{cm23park}.
Moreover, QNN-based models are generally able to achieve fast convergence and high scalability, not only in IoT devices but also in various computing platforms~\cite{cm23park}.}

\rev{However, QNN-based models can be more difficult to be trained comparing to the training of conventional DNN-based models because of \textit{barren plateaus}, where it is defined as the nonlinearity index depending on the amounts of entanglements in quantum circuits~\cite{mcclean2018barren}. Therefore, it is essentially required to track the barren plateaus for measuring the stability of QNN software}. 

\rev{In the perspective of testing, analysis, and code optimization of QNN software}, it should be helpful to obtain high-accurate \rev{QNN execution} results in a fast way, even \rev{if software engineers} are not familiar with the concepts of quantum \rev{computing}.
\rev{Therefore, it} is essentially required \rev{to design and implement} a new software \rev{test, analysis, and code optimization (TACO)} tool for QNN software, which should visually identify barren plateau situations \rev{for the explainability and trainability of QNN}. In addition, it should provide feedback that is useful for software testing\rev{, analysis, and code optimization}. Based on this \rev{TACO tool for QNN software}, software engineers are able to re-organize \rev{and optimize} their own QNN software codes. Here, the concept of human-computer interaction (HCI) \rev{can be} involved because the human (software engineers) can re-organize \rev{and optimize} the codes for the computer (\rev{IoT devices}); in turn, the computer \rev{can} visualize the barren plateau situations to \rev{the} software engineers. \rev{These} barren plateau \rev{situations generally} occur in QNN-based models due to the operations of \textit{i)} quantum controlled gates and \textit{ii)} the superposition of qubits.
In addition, according to the entanglement among multiple qubits, QNN-based models are able to improve their own performances. 
These entangled quantum states are advantageous in terms of capacity, whereas multi-qubit interference increases while many quantum controlled gates and qubits are simultaneously operated. 
This introduces the \rev{situation} where QNN-based model training optimization is impossible \rev{due to} barren plateaus. 
Therefore, it is critical to design high-accurate QNN-based models for improving learning training performance. 
Among significant research contributions to deal with the barren plateau problem, it has been verified that the barren plateau problem is associated with the structure of QNN-based models~\cite{mcclean2018barren}. Therefore, controlling the system size in the structure of QNN-based models enhances the performance of quantum learning training. 
During this procedure, general software engineers will be in trouble in terms of \textit{i)} understanding quantum mechanics, quantum computing, and quantum neural networks (including the definition of barren plateaus) and \textit{ii)} \rev{the} identification of optimal system size in the structure of QNN-based models in order to reduce quantum entanglement interference. 

Motivated by this, a novel software testing \rev{and analysis} tool, \textit{i.e.}, TACO is designed which enables dynamic run-time software testing \rev{and analysis} in \rev{advanced IoT systems}. As depicted in Fig.~\ref{fig:sec2}, quantum-based advanced IoT systems \rev{are} capable to perform various tasks, including \rev{surveillance for IoT sensors, recognition for mobile devices, and sensing/processing for wearable devices.} During this process, our software testing \rev{and analysis} tool, TACO, is able to provide real-time updates on the information that is trained by advanced IoT systems, thereby enabling the tool to verify the accuracy of the perception during run-time execution.

\rev{The main contributions and objectives of this paper can be categorized and summarized as follows.
\begin{itemize}
    \item Firstly, our proposed TACO is the world-first software testing, analysis, and code optimization tool of QNN-based models in advanced IoT systems. The main objective of TACO is for pursuing high-accurate and stabilized performance maintenance in the QNN-based models by controlling the degree of entanglements. 
    \item Moreover, the additional objective of TACO is to  provide intuitive approaches such as visualization and HCI-based feedback for general software engineers during QNN-based model design and implementation. 
    \item Lastly, this paper provides the various extension directions of TACO those are considerable for emerging applications. 
\end{itemize}
}

\rev{The rest of this article is organized as follows. Sec.~\ref{sec:quantumsupremacy} introduces quantum supremacy and the limitations. In addition, Sec.~\ref{sec:quest} presents the details of our proposed TACO for QNN software testing and visualization. Moreover, Sec.~\ref{sec:open} outlines relevant corresponding open discussions, and lastly, Sec.~\ref{sec:conclusions} concludes this paper.}

\section{Quantum Supremacy and Limitations}\label{sec:quantumsupremacy}
\rev{This section firstly introduces the foundations of QNN (refer to Sec.~\ref{sec:qnn}), and then discusses its advantages (refer to Sec.~\ref{sec:qnn-pros}) and one of major disadvantages which is called barren plateau (refer to Sec.~\ref{sec:bp}), respectively.} 

\subsection{Quantum Neural Networks}\label{sec:qnn}
The conventional neural network can be mathematically interpreted as the sequential combination of multiple \textit{i)} linear transformation and \textit{ii)} nonlinear \rev{transformation}.
The structure of QNN-based models is basically similar to the structure of conventional neural networks. Note that one of the major differences between the two neural networks is that the training/inference computation of QNN-based models is performed over Hilbert spaces whereas the computation of conventional neural networks is performed over real number spaces. Therefore, the QNN-based models optimize their objective functions in quantum states over Hilbert spaces.
The detailed structural components of QNN-based models, called \textit{quantum gates}, are \textit{i)} quantum rotation gates for quantum state transformation over Bloch sphere and \textit{ii)} quantum-controlled gates for entanglement generation with other qubits.
With the utilization of these quantum gates, the structure of \rev{QNN-based models} consists of \textit{i)} \textit{state encoding} (encoding the classical data over real number spaces to the quantum states over Hilbert spaces, \textit{i.e.}, converting classical $0$ or $1$ bits into qubits), \textit{ii)} \textit{parameterized quantum circuit \rev{(PQC)}} (controlling the rotations and entanglements for input quantum states under the utilization of quantum gates), and \textit{iii)} \textit{measurement} (decoding the quantum states over Hilbert spaces into the classical data over real number spaces, \textit{i.e.}, converting qubits into classical $0$ or $1$ bits, during training optimization). 
After this structural computation, the quantum states in the QNN-based models can be observed; and this is conducted for minimizing its own \rev{cost} function.
The procedures \textit{i--ii)}, \textit{i.e.}, \textit{state encoding} and \textit{parameterized quantum circuit}, are linear transformations, and the procedure \textit{iii)}, \textit{i.e.}, \textit{measurement}, is a unique non-linear \rev{transformation}~\cite{you2021exponentially}.

This QNN-based model is advantageous compared to conventional neural networks in terms of data processing acceleration and low computational complexity. However, this QNN-based model has disadvantages as well; and one of the disadvantages is the barren plateaus \rev{situation} which can make the loss gradient of QNN-based models vanish \rev{due to} entanglement. More details about this barren plateaus problem will be discussed later (refer to Sec.~\ref{sec:bp}).

\begin{figure*}[h!]
\centering
\includegraphics[width=0.68\linewidth]{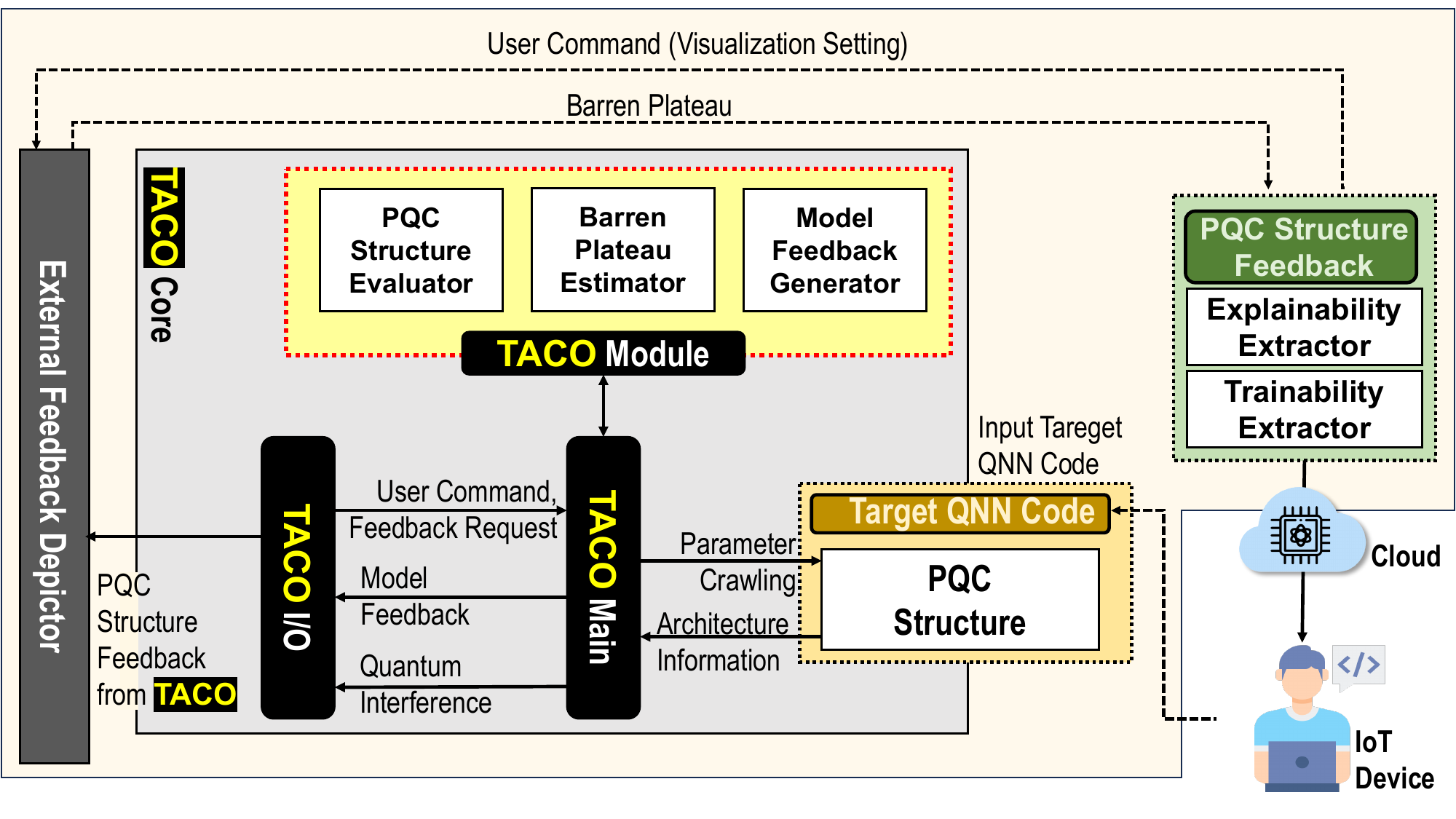}
\caption{Overall Process of TACO, \textit{i.e.}, Dynamic Run-Time Quantum \rev{Software \textit{Testing, Analysis, and Code Optimization}}.}
\label{fig:system_archi}
\end{figure*}

\subsection{\rev{Advantages in QNN}}\label{sec:qnn-pros}

\rev{The advantages of QNN for can be listed as follows.}

\begin{itemize}

    \item \rev{\BfPara{High Scalability} As theoretically well-justified in~\cite{cikm23baek}, 
    QNN-based models can expand the output dimensions to an exponential scale by incorporating Pauli-$Z$ measurement and basis measurement, surpassing the limitations imposed by the limited number of qubits in the noisy intermediate scale quantum (NISQ) era.}
    
    \item \rev{\BfPara{Fewer Parameter Utilization} QNN-based models have the ability to achieve similar performance to conventional neural networks with fewer parameters, primarily due to superposition and entanglement. Firstly, the superposition allows the quantum models to exist in multiple states at once simultaneously, and thus, it expands the representational dimension. Moreover, the entanglement provides solid relationship among quantum states using controlled-NOT (CNOT) gates, and thus, it realizes nonlinear representation with fewer parameters.} 

    \item \rev{\BfPara{Fast Convergence} In QNN, the \textit{parameter shift rule} is used during training instead of the backpropagation in conventional neural network training. The \textit{parameter shift rule} provides simple and direct approaches, and thus, QNN-based model training can be accelerated further than conventional neural network training.} 
    
\end{itemize}

\begin{figure*}[t!]
\centering
\includegraphics[width=0.59\linewidth]{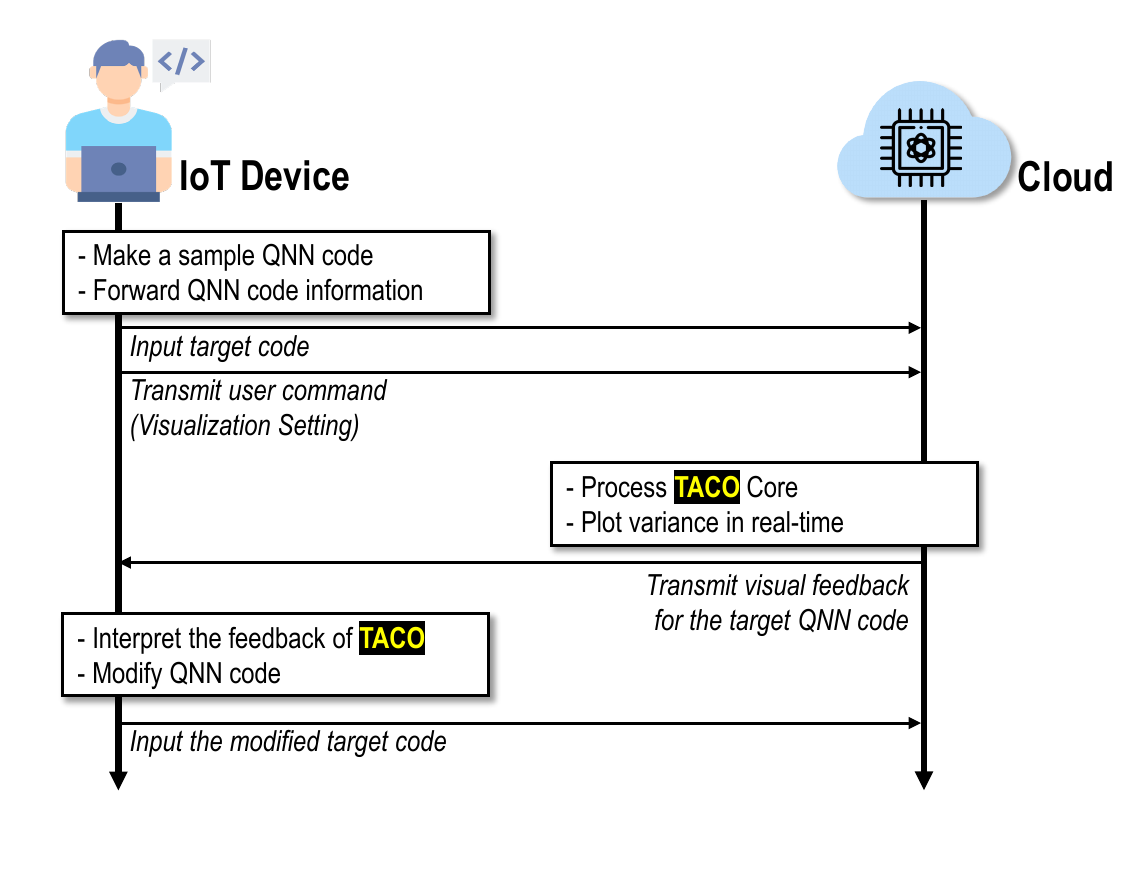}
\caption{\rev{Flowchart of TACO}}
\label{fig:flowchart}
\end{figure*}

\subsection{Barren Plateaus}\label{sec:bp}
The barren plateau is one of the well-known harmful \rev{problems} for \rev{QNN-based learning model} training optimization where the loss function gradients are exponentially and dramatically \rev{degradated} due to entanglement~\cite{mcclean2018barren}; and this can lead to local minima during \rev{cost} function minimization.
Moreover, \rev{recent remarkable research results} confirm that \rev{a lot of} barren plateaus can occur during QNN-based model training optimization~\cite{you2021exponentially}.
\rev{For more details}, the number of barren plateaus is proportioned to the number of qubits exponentially, and \rev{this means} the occurrence of barren plateaus depends on QNN-based model design and implementation. 
\rev{Furthermore}, it is \rev{hard} to escape from the barren plateaus~\cite{mcclean2018barren}.
In order to tackle \rev{this situation}, various approaches have been \rev{suggested}. However, these research results fundamentally require deep-dive understanding in terms of quantum mechanics, quantum computing, and quantum optimization.
To resolve this \rev{problem} from the viewpoints of \rev{deep} learning software engineers, it is obvious that a novel software development \rev{and analysis} tool is required which can efficiently inform the barren plateau occurrences during QNN-based model training \rev{optimization}.

\section{TACO for QNN-based Models}\label{sec:quest}

\subsection{Design}
The \rev{details of} three fundamental design rationales of TACO are as follows.

\subsubsection{Barren Plateaus Problem Tracking} 
For the design of high-accurate \rev{QNN-based model} software, TACO informs the barren plateaus situation occurrence and its corresponding performance degradation to quantum machine learning software developers and engineers. According to this beneficial functionality of TACO, the QNN software engineers are able to significantly reduce the degraded QNN-based model training performances. As a result, software engineers can achieve robust and resilient QNN-based model training \rev{optimization}. 

\subsubsection{Dynamic Run-Time Software Testing}
It is interesting to know that the software testing \rev{and analysis} of QNN-based model software codes is not able to be conducted using static analysis. This is because the static analysis conducts the software testing for the integrity of QNN software with original codes. The reason is that the QNN software codes should be tested and verified while qubits inputs are fed to QNN-based models where the qubits exist in probabilistic quantum states before the measurement. Hence, it is essential to conduct the software testing for the given \rev{QNN-based model software} codes using dynamic run-time analysis which tests the integrity of \rev{QNN} software codes during their run-time executions.

\subsubsection{HCI-based Visualization} 
For identifying the barren plateaus situations those are harmful for the \rev{high-accurate} execution of QNN-based models, TACO observes the dynamics of barren plateaus (\textit{i.e.}, variances of gradients) while the qubit state changes over time. Therefore, the visualization of the barren plateau dynamics over time is fundamentally required for intuitive understanding to QNN software engineers. Then, they can easily understand the performance of the given QNN-based models, even if the software engineers are not familiar with the basic knowledge of quantum mechanics, quantum computing, and quantum machine learning. Lastly, these visualization results can provide sufficient information to software engineers, and then, they can provide feedback information to TACO via HCI-based visual feedback. 

\subsection{Implementation}

\begin{figure*}
    \centering
    \includegraphics[width=1.49\columnwidth]{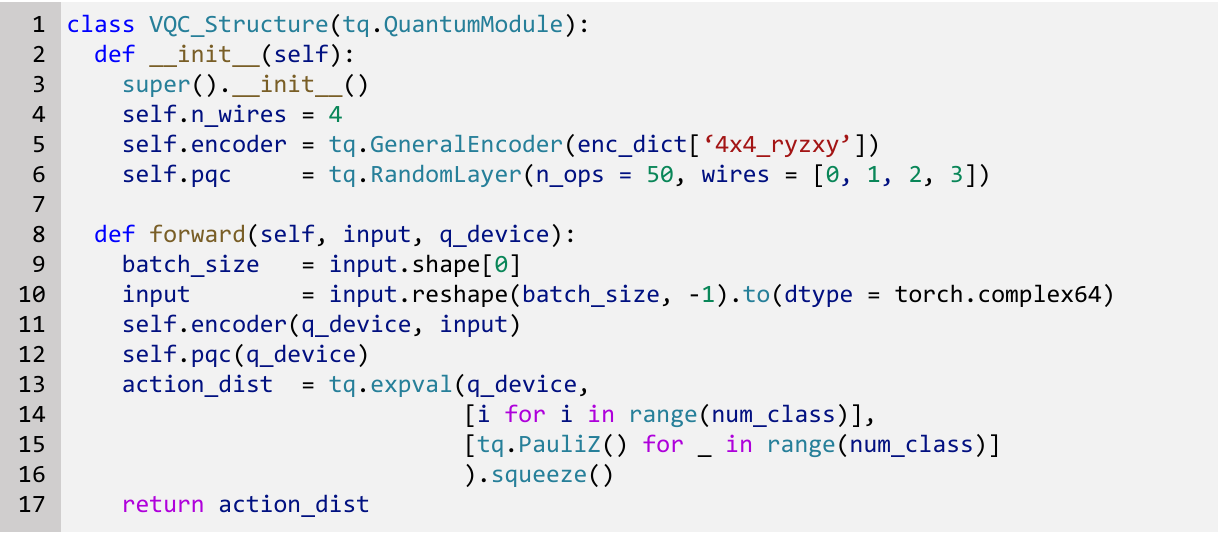}
    \caption{The Example Code for \textsf{VQC Structure}.}
    \label{fig:VQC_Structure}
\end{figure*}

\begin{figure}
    \centering
    \includegraphics[width=0.95\columnwidth]{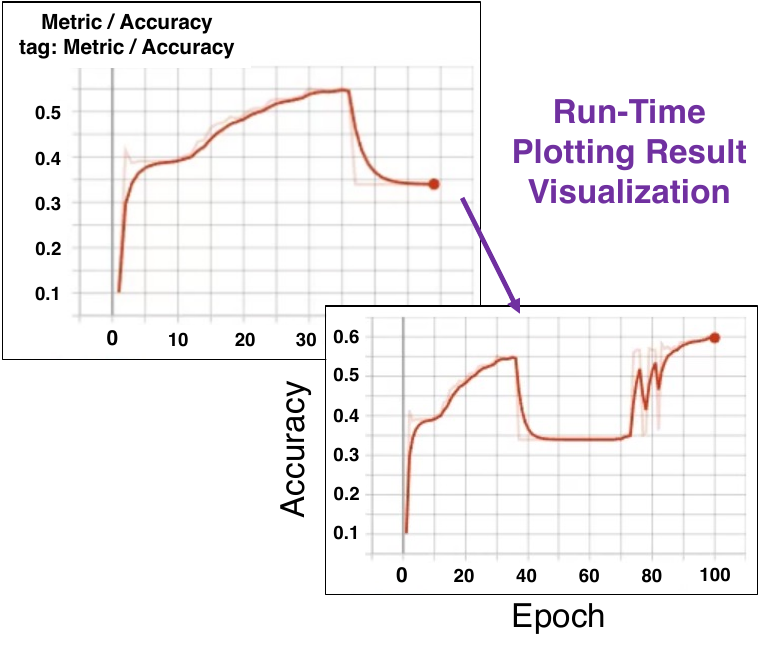}
    \caption{Test Accuracy.}
    \label{fig:Test_accuracy}
\end{figure}

\begin{figure}
    \centering
    \includegraphics[width=0.95\linewidth]{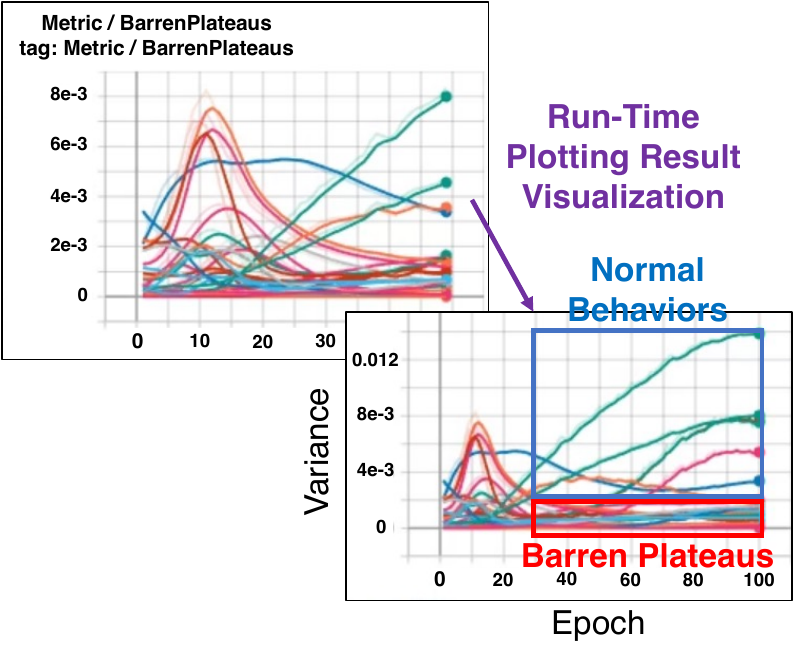}
    \caption{Barren Plateaus.}
    \label{fig:performance}
\end{figure}




The overall software \rev{architectures} and their corresponding implementation of TACO are illustrated in Fig.~\ref{fig:system_archi}, and the detailed \rev{descriptions} are as follows.

\begin{itemize}
    \item \textsf{VQC Structure}: This is a \rev{variational} quantum circuit \rev{(VQC)} and this is the user input by software engineers. In our implementation by the \rev{software}engineers, the considering \rev{QNN-based model software} code is implemented by \texttt{torch-quantum}. 
    In the QNN-based model software code, the quantum rotation gates and controlled gates are organized and implemented by the random layers in \texttt{torch-quantum}. The example implementation of \textsf{VQC Structure} is presented in Fig.~\ref{fig:VQC_Structure}. 
    For \textit{(Lines 4--6)} in Fig.~\ref{fig:VQC_Structure}, the defined class of \textsf{VQC Structure} is with \texttt{n\_wires} which stands for the number of input qubits. \rev{Additionally}, this has \texttt{encoder} which converts classical bits into qubits. Moreover, \texttt{pqc} exists in this \textsf{VQC Structure} class and this consists of multiple quantum rotations and controlled gates. 
    For \textit{(Line 8--17)} in Fig.~\ref{fig:VQC_Structure}, the function called \texttt{forward} is defined when the object of the QNN-based model is generated. The input of this object feeds into \texttt{encoder} and \texttt{pqc}, sequentially. 
    After conducting this procedure, \texttt{tq.PauliZ()} (defined in \texttt{tq.expval()}) is able to compute projections over $z$-axis. Finally, the output result of a measurement can be obtained, and this is the computed and approximated action of our trained QNN-based models. 
    \item \textsf{TACO Engine}: Our TACO engine in Fig.~\ref{fig:system_archi} consists of following three modules, \textit{i.e.}, \textsf{VQC Structure Extractor}, \textsf{Barren Plateau Estimator}, and \textsf{Model Feedback Generator}. 
    \begin{itemize}
        \item \textsf{VQC Structure Extractor}: This calls parameter information via \texttt{named\_parameters()} in \texttt{torch-quantum}; and the iterative gradient derivation calculation is \rev{operated} via \texttt{backward()}.  
        \item \textsf{Barren Plateau Estimator}: This estimates the barren plateau occurrences and values where the value stands for the variance of gradients in the quantum gates of QNN-based models. 
        For more details, we can \rev{track} the computation falls into a barren plateau when the variance of the gradient \rev{suddenly drops}. The implementation of this variance of the gradient can be by \texttt{var()} which is one of the statistics functions in \texttt{torch-quantum}. 
        \item \textsf{Model Feedback Generator}: This generator handles the case where the barren plateau \rev{situations} (\textit{i.e.}, the variance of gradients) \rev{happens} in all quantum gates. In addition, this generates text messages which include \texttt{epoch}, \texttt{parameter index}, \texttt{parameter type}, and \texttt{barren plateaus value}.       
    \end{itemize}
    \item \textsf{TACO I/O}: It manages the inputs and outputs of TACO. It sends the data of VQC structure feedback for visualization to \textsf{External Visualization Engine} (\textit{i.e.}, \textsf{tensorboard} in our implementation). 
    \item \textsf{External Visualization Engine}:  
    This engine is for recording the data from TACO and \rev{presenting} to our considering \textsf{tensorboard}. 
    The training data can be plotted using this engine. This \rev{conducts} the visualization of `train loss', `test accuracy', and `barren plateaus' values.    
    Then, QNN software engineers are able to identify training data via \textsf{tensorboard} and also check which local parameters are in barren plateau situations. 
    
\end{itemize}

\rev{Based on this implementation, the operational flowchart can be illustrated as Fig.~\ref{fig:flowchart}. Firstly, IoT devices make a sample QNN code and forward the QNN code information to their associated cloud, i.e., target code input. Additionally, IoT devices transmit the user command for visualization setting. Then, the could processes TACO Core and plots variance in real-time for barren plateaus tracking. Then, the results will be transmitted to its associated IoT devices for the visual feedback pf the target QNN code. After receiving the feedback at IoT devices, they interpret the feedback of TACO and modify their QNN codes based on that. Finally, the IoT devices deliver the modified target QNN code input to their associated cloud.}

\subsection{Visualization}
The \rev{run-time} visualization results for dynamic run-time software testing \rev{and analysis} using TACO are illustrated in Fig.~\ref{fig:Test_accuracy} and Fig.~\ref{fig:performance}, where Fig.~\ref{fig:Test_accuracy} shows the test accuracy over time during QNN-based model training \rev{optimization} and Fig.~\ref{fig:performance} shows the variance\rev{s} over time in quantum rotation gates (\textit{i.e.}, $R_X$, $R_Y$, and $R_Z$) in VQC at every epoch. If the variance\rev{s} \rev{suddenly drop}, it is obvious that the corresponding quantum gates fall in barren plateaus situations. As \rev{shown} in Fig.~\ref{fig:Test_accuracy}, the test accuracy drops to approximately $35$\% from $35$ to $75$ epochs, where the number of quantum rotation gates those are in barren plateau situations is the highest as \rev{demonstrated} in Fig.~\ref{fig:performance}. Accordingly, it can be \rev{observed} that there \rev{exists} a negative relationship between the performance of QNN-based models and barren plateau situations.
Moreover, Fig.~\ref{fig:performance} presents run-time performance evaluation results plotting over time with visualization. Therefore, it is able to provide intuitive understanding for software engineers even in cases where they are not familiar with quantum mechanics, quantum computing, and quantum \rev{optimization}.
Finally, the proposed TACO can identify which quantum gates fall \rev{into} barren plateaus situations with text messages.

In summary, software engineers are able to monitor and observe QNN-based model training performances using TACO. If the performance of current QNN-based model training \rev{optimization} is less than expectation, the software engineers can identify the problematic quantum rotation gates using 'Model Feedback' in the proposed TACO. Thanks to this advantage of this TACO, the software engineers can adequately control the parameters of the problematic quantum rotation gates in order to improve QNN-based model training performances.

\section{Open Discussions}\label{sec:open}
A dynamic run-time software testing \rev{and analysis} tool, called TACO, is proposed and it is \rev{for} verifying the trainability and explainability of \rev{QNN-based model} software for \rev{advanced IoT systems}. In order to realize the QNN functionalities in \rev{advanced IoT systems}, it is required to address open issues in terms of the usability of quantum computing technologies, \textit{i.e.}, \textit{i)} quantum computer miniaturization, \textit{ii)} limitations in the NISQ era, and \textit{iii)} commercialization. 

\subsection{Quantum Computer Miniaturization}
\rev{Recently, most of the research contributions in the areas of quantum computing, quantum algorithms, and QNN are aiming to improve computation performance in terms of speed and computing resources. Therefore, industry companies such as \textit{IBM} and \textit{Google} have produced large quantum computing platforms. As a result, it} has been widely known that quantum computers are heavy and must be stored in cryogenic environments. Therefore, people may have questions about the feasibility of quantum computer utilization in \rev{IoT devices} for \rev{QNN-based model} utilization. However, there are already triumphant attempts at building miniaturized quantum computers. For example, \textit{Alpine Quantum Technologies} has developed a commercial 19-inch rack-mounted 20-qubit quantum computer; and demonstrating that it is possible to lessen the size of quantum computers~\cite{blatt_monz_zoller_2022}. For more details, it only requires two fully custom racks, runs in a room temperature environment, and consumes less than 2\,kW of power. Based on the pace of progress in quantum computers, the potential of QNN utilization for reliable \rev{IoT systems} is significant and it is obvious that the quantum computer miniaturization technology development is enough to be realized. \rev{Moreover, \textit{Quantum Brilliance} developed a 5-qubit miniature quantum computing platforms with the size of a desktop mainframe. Lastly, \textit{SpinQ} created a portable 2-qubit quantum computer called \textit{Gemini mini}.}

\subsection{Limitation in the NISQ Era}
Based on the utilization of quantum computing, it is possible to realize accurate and accelerated massive \rev{large-scale} data processing. However, according to the limitations of hardware technologies in quantum computing, the number of qubits that can be utilized \rev{for real-time computation} is strictly limited in the noisy intermediate-scale quantum (NISQ) era.
In this case, quantum errors occur during the computation with quantum gates and circuits, when a relatively larger number of qubits is utilized. This is definitely harmful to \rev{advanced IoT related} decision-making if the decision-making is made by QNN-based models. 
In order to deal with this problem, significant \rev{and novel} research results have been proposed for \rev{taking care of} quantum errors in \rev{advanced IoT systems}. 
For example, the proposed algorithm in~\cite{baek2022sqcnn3d} conducts the classification tasks during 3D point cloud data processing for \rev{IoT-capable} autonomous driving with a small number of qubits. It is also confirmed that this algorithm outperforms other learning-based methods. 
\rev{Moreover}, the proposed algorithms in~\cite{10012051,iotj23park} are for quantum actor-critic reinforcement learning in order to control multiple autonomous mobile robots and multiple unmanned aerial vehicles, respectively. \rev{These algorithms} outperform other learning-based methods even though \rev{they} utilize a small number of qubits. 
\rev{Furthermore}, many experts in QNN and quantum computing predict that several hundreds of qubits can be utilized where the qubits are in low-noise situations~\cite{preskill2018quantum}.
For \rev{example}, \rev{academia and industry} experts expect that \textit{IBM} will design its own quantum computer which can utilize $4,000$ qubits until 2025~\cite{commercialization_2022}. 
\rev{Lastly}, \textit{Baidu} designed and implemented its own 10-qubit quantum computer that is \rev{named to} \textit{Qian-Shi}; and insists that the quantum computer can deal with many real-world \rev{on-hand} problems~\cite{commercialization_2022}.

\subsection{Commercialization}
In order to deploy QNN and quantum computing technologies in \rev{advanced IoT systems}, the mass production and commercialization of quantum computers should be realized \rev{and established}. 
\textit{Shenzhen SpinQ Technology}, which is \rev{one of} quantum computing startup\rev{s} in China, sells portable quantum computers with \rev{the} prices from $5,000$ to $9,000$ US dollars~\cite{price_2022}. 
\rev{Furthermore}, it is \rev{clear} that the prices can be \rev{dramatically} decreased for mass production and commercialization. 


\section{Conclusions and Future Work}\label{sec:conclusions}
This paper \rev{introduces} a novel dynamic run-time software testing\rev{, analysis, and code optimization (TACO)} tool for \rev{QNN-based model} software, which visually presents gradient variances in order to identify whether barren plateaus occur or not for advanced IoT systems software. 
This \rev{TACO} tool is obviously useful for software engineers because it can intuitively guide them in order to design and implement high-accurate QNN-based models for \rev{advanced IoT} applications even if they are not familiar with quantum mechanics and quantum computing.
Moreover, the proposed TACO is also capable for visual feedback because software engineers recognize barren plateaus using visualization via tensorboard; and then, they modify QNN-based model structures based on that.

\rev{As future work directions, our proposed TACO can be extended in various ways as follows.
\begin{itemize}
    \item Firstly, the current version of TACO focuses on barren plateaus whereas QNN-based models are with various performance-dependent factors in the \textit{state encoder} and \textit{measurement} of QNN. Therefore, the extension of TACO should be able to track and test the factors. 
    \item Moreover, our proposed TACO can be extended for distributed learning architectures such as quantum slit learning~\cite{cikm23park}. In quantum split learning, single QNN is divided into two parts and they are located in separated computers~\cite{icoin20jeon}. For the separated parts, one can be with the state encoder of QNN; and the other one can be with the PQC and measurement of QNN. Then the first part should create dummy codes for PQC and measurement in order to generate the complete QNN software for conducting TACO-based testing and analysis. Similarly, the other part should create dummy codes for state encoder in order to generate the complete QNN software for conducting TACO. This approach is fundamentally based on automated code generation and various related methods should be studied under the consideration of quantum computing characteristics. 
\end{itemize}
}
\bibliographystyle{IEEEtran}

\begin{thebibliography}{10}
\providecommand{\url}[1]{#1}
\csname url@samestyle\endcsname
\providecommand{\newblock}{\relax}
\providecommand{\bibinfo}[2]{#2}
\providecommand{\BIBentrySTDinterwordspacing}{\spaceskip=0pt\relax}
\providecommand{\BIBentryALTinterwordstretchfactor}{4}
\providecommand{\BIBentryALTinterwordspacing}{\spaceskip=\fontdimen2\font plus
\BIBentryALTinterwordstretchfactor\fontdimen3\font minus \fontdimen4\font\relax}
\providecommand{\BIBforeignlanguage}[2]{{%
\expandafter\ifx\csname l@#1\endcsname\relax
\typeout{** WARNING: IEEEtran.bst: No hyphenation pattern has been}%
\typeout{** loaded for the language `#1'. Using the pattern for}%
\typeout{** the default language instead.}%
\else
\language=\csname l@#1\endcsname
\fi
#2}}
\providecommand{\BIBdecl}{\relax}
\BIBdecl

\bibitem{ic2309park}
S.~Park, H.~Feng, C.~Park, Y.~Lee, S.~Jung, and J.~Kim, ``{EQuaTE}: Efficient quantum train engine for runtime dynamic analysis and visual feedback in autonomous driving,'' \emph{IEEE Internet Computing}, vol.~27, no.~5, pp. 24--31, September-October 2023.

\bibitem{tvt2108jung}
S.~Jung, J.~Kim, M.~Levorato, C.~Cordeiro, and J.-H. Kim, ``Infrastructure-assisted on-driving experience sharing for millimeter-wave connected vehicles,'' \emph{IEEE Transactions on Vehicular Technology}, vol.~70, no.~8, pp. 7307--7321, August 2021.

\bibitem{cm23park}
S.~Park, J.~P. Kim, C.~Park, S.~Jung, and J.~Kim, ``Quantum multi-agent reinforcement learning for autonomous mobility cooperation,'' \emph{IEEE Communications Magazine}, 2023 (Early Access).

\bibitem{mcclean2018barren}
J.~R. McClean, S.~Boixo, V.~N. Smelyanskiy, R.~Babbush, and H.~Neven, ``Barren plateaus in quantum neural network training landscapes,'' \emph{Nature Communications}, vol.~9, no.~1, p. 4812, November 2018.

\bibitem{you2021exponentially}
X.~You and X.~Wu, ``Exponentially many local minima in quantum neural networks,'' in \emph{Proc. Int'l Conf. Machine Learning (ICML)}, Virtual, July 2021.

\bibitem{cikm23baek}
H.~Baek, S.~Park, and J.~Kim, ``Logarithmic dimension reduction for quantum neural networks,'' in \emph{Proc. ACM Conf. Information and Knowledge Management (CIKM)}, Birmingham, U.K., October 2023.

\bibitem{blatt_monz_zoller_2022}
\BIBentryALTinterwordspacing
R.~Blatt, T.~Monz, and P.~Zoller, ``The world's leading 19'' rack-mounted quantum computer,'' May 2022. [Online]. Available: \url{https://www.aqt.eu/pine-system-19-rack-mounted-quantum-computer/}
\BIBentrySTDinterwordspacing

\bibitem{baek2022sqcnn3d}
H.~Baek, W.~J. Yun, and J.~Kim, ``{3D} scalable quantum convolutional neural networks for point cloud data processing in classification applications,'' in \emph{Proc. AAAI Workshop on AI to Accelerate Science and Engineering (AI2ASE)}, Washington, DC, USA, February 2023.

\bibitem{10012051}
W.~J. Yun, J.~P. Kim, S.~Jung, J.-H. Kim, and J.~Kim, ``Quantum multiagent actor-critic neural networks for {Internet}-connected multirobot coordination in smart factory management,'' \emph{IEEE Internet of Things Journal}, vol.~10, no.~11, pp. 9942--9952, June 2023.

\bibitem{iotj23park}
C.~Park, W.~J. Yun, J.~P. Kim, T.~K. Rodrigues, S.~Park, S.~Jung, and J.~Kim, ``Quantum multi-agent actor-critic networks for cooperative mobile access in multi-{UAV} systems,'' \emph{IEEE Internet of Things Journal}, vol.~10, no.~22, pp. 20\,033--20\,048, November 2023.

\bibitem{preskill2018quantum}
J.~Preskill, ``Quantum computing in the {NISQ} era and beyond,'' \emph{Quantum}, vol.~2, p.~79, August 2018.

\bibitem{commercialization_2022}
\BIBentryALTinterwordspacing
J.~Bourne, ``Commercial quantum computer disruption on the horizon,'' August 2022. [Online]. Available: \url{https://www.insiderintelligence.com/content/commercial-quantum-computer-disruption-on-horizon}
\BIBentrySTDinterwordspacing

\bibitem{price_2022}
\BIBentryALTinterwordspacing
H.~Corrigan, ``You can buy a portable quantum computer for under \$9{K},'' December 2022. [Online]. Available: \url{https://www.pcgamer.com/you-can-buy-a-portable-quantum-computer-for-under-dollar9k/}
\BIBentrySTDinterwordspacing

\bibitem{cikm23park}
S.~Park, H.~Baek, and J.~Kim, ``Quantum split learning for privacy-preserving information management,'' in \emph{Proc. ACM Conf. Information and Knowledge Management (CIKM)}, Birmingham, U.K., October 2023.

\bibitem{icoin20jeon}
J.~Jeon and J.~Kim, ``Privacy-sensitive parallel split learning,'' in \emph{Proc. IEEE Int'l Conf. Information Networking (ICOIN)}, Barcelona, Spain, January 2020.

\end{thebibliography}

\begin{IEEEbiographynophoto}{Dr. Soohyun Park}
has been a postdoctoral scholar at the Department of Electrical and Computer Engineering, Korea University, Seoul, Republic of Korea, since September 2023. She received the Ph.D. degree in electrical and computer engineering from Korea University, Seoul, Republic of Korea, in August 2023. She received the B.S. degree in computer science and engineering from Chung-Ang University, Seoul, Republic of Korea, in February 2019. 
She was a recipient of the IEEE Vehicular Technology Society (VTS) Seoul Chapter Awards (2019, 2023), IEEE Seoul Section Student Paper Content Awards (2020, 2023), and \textit{ICT Express (Elsevier)} Best Reviewer Award (2021).
\end{IEEEbiographynophoto}

\vspace{-10mm}

\begin{IEEEbiographynophoto}{Prof. Joongheon Kim}
(M'06--SM'18) has been with Korea University, Seoul, Korea, since 2019, and he is currently an associate professor. He received the Ph.D. degree in computer science from the University of Southern California (USC), Los Angeles, CA, USA, in 2014. He serves as an editor for \textsc{IEEE Transactions on Vehicular Technology} and \textsc{IEEE Internet of Things Journal}. He was a recipient of Annenberg Graduate Fellowship with his Ph.D. admission from USC (2009), Intel Corporation Next Generation and Standards (NGS) Division Recognition Award (2015), \textsc{IEEE Systems Journal} Best Paper Award (2020), IEEE ComSoc Multimedia Communications Technical Committee (MMTC) Outstanding Young Researcher Award (2020), and IEEE ComSoc MMTC Best Journal Paper Award (2021). He also received IEEE ICOIN Best Paper Award (2021), IEEE Vehicular Technology Society (VTS) Seoul Chapter Awards (2019, 2023), and IEEE ICTC Best Paper Award (2022). 
\end{IEEEbiographynophoto}

\end{document}